# What Makes Audio Event Detection Harder than Classification?

Huy Phan*†, Philipp Koch*, Fabrice Katzberg*, Marco Maass*, Radoslaw Mazur*,
Ian McLoughlin‡ and Alfred Mertins*
*Institute for Signal Processing, University of Lübeck, Germany
†Graduate School for Computing in Medicine and Life Sciences, University of Lübeck, Germany
‡School of Computing, University of Kent, UK
{phan, koch, katzberg, maass, mazur, mertins}@isip.uni-luebeck.de, ivm@kent.ac.uk

*Abstract*—There is a common observation that audio event classification is easier to deal with than detection. So far, this observation has been accepted as a fact and we lack of a careful analysis. In this paper, we reason the rationale behind this fact and, more importantly, leverage them to benefit the audio event detection task. We present an improved detection pipeline in which a verification step is appended to augment a detection system. This step employs a high-quality event classifier to postprocess the benign event hypotheses outputted by the detection system and reject false alarms. To demonstrate the effectiveness of the proposed pipeline, we implement and pair up different event detectors based on the most common detection schemes and various event classifiers, ranging from the standard bag-of-words model to the state-of-the-art bank-of-regressors one. Experimental results on the ITC-Irst dataset show significant improvements to detection performance. More importantly, these improvements are consistent for all detector-classifier combinations.

## I. INTRODUCTION

Audio event classification and detection (AEC/D) have been an active field of research in recent years [1], [2], [3]. So far, beside a majority of works focusing on the improving overall performance in terms of accuracy [2], [1], [4], [5], many other aspects have also been studied, including noise robustness [6], [7], [8], overlapping event handling [9], [10], [11], [12], early event detection [13], multi-channel fusion [14], as well as generic representation [15]. However, little attention has been paid to the important aspect of event detection systems on continuous streams: false positive reduction. False positives, i.e., event instances that are spuriously detected by a detection system, and subsequently draw attention to them, are arguably one of the most important problems faced by different applications like ambient intelligence and surveillance. To the best knowledge of the authors, this is the first work explicitly addressing this problem.

Previous research on audio event detection can be roughly grouped into three different schemes. The first one is the detection-by-classification where event classification models, for example Support Vector Machines (SVMs), are learned and then applied on test audio signals in a sliding-window fashion [16], [17], [5]. The second scheme relies on automatic speech recognition (ASR) frameworks where the states of frame-wise features are modeled by Gaussian Mixture Models (GMMs) followed by Hidden Markov Models (HMMs) to model the distributions of the feature sequences given the state sequences [18], [19], [20]. On testing, the target event is recognized by maximizing the posterior probability on a local feature vector sequence of a test audio signal. The third scheme is based on the recently proposed regression approach [4], [13]. A regressor based on random regression forests is learned for each target event category to model the relative positions of the audio segments with respect to the event onsets and offsets. On testing, the learned regressor is used to estimate the positions of the event onsets and offsets (i.e. their boundaries) in a test audio signal.

The goal of false positive reduction is obviously achievable by improving the overall performance towards an oracle system which makes no mistakes. This has inherently been the main focus of many works since the task was introduced. Nevertheless, our purpose is different from them in essence. We aim to reduce the false positives of detection systems given the state-of-the-art performance which is far from perfect in practice. Our proposed detection pipeline is inspired by investigating the performance gap between a detection system on continuous streams and a classification system on isolated events. More specifically, the classification task is much easier to deal with and usually enjoys higher overall accuracy than that of the event detection task. Based on this observation, instead of making hard detection decisions early on, it is reasonable to inject a verification step at the end of the detection pipeline, where a trained classifier is used to classify the detected hypotheses and reject those with mismatched labels. Unlike the common detection-by-classification scheme [16], [17], [1], this can be considered as a novel scheme to utilize a trained event classifier for the detection task.

## II. THE IMPROVED DETECTION PIPELINE FOR FALSE POSITIVE REDUCTION

### A. Why is audio event detection harder than classification?

There is a fact that the classification task is much easier than the detection one. For example, on the ITC-Irst dataset [2] used in this work accuracies of 98.9% and 93.1% are obtainable for the classification and detection tasks, respectively (cf. Section IV). This observation is also well-known in the CLEAR 2006 challenge [2]. This can be explained by the fundamental difference between the tasks. The reason is two-fold. First and more obviously, the detection task needs to discriminate not only the event categories of interest (as in the classification task) but also the target event categories from highly rich background sounds. Second, for the classification task, we have access to the global context of the events while, in the detection task, we do not know in advance boundaries of the events and usually need to rely on unreliable local audio features for inference.

Furthermore, this fundamental dissimilarity results in a drawback of event detectors which are based on the common detection-by-classification scheme, such as those in [17]. These detectors attempted to build strong event classification models and subsequently employ them to detect events in continuous streams with a sliding window. Since the classifiers are trained on complete events, they expect to be presented with complete events to guarantee a good performance. However, due to high intra- and extra-class temporal variations of audio events, it is almost impossible to decide a good-for-all window length that exactly captures complete events. This results in mismatch between training and testing data, which significantly deteriorates the accuracy of the classification models and subsequently the accuracy of detection systems. Although one can circumvent this issue by training classifiers on equal-sized segments of the events, such as those in [5], [16], the problem remains unsolved. By dividing the events into equal-sized segments, one has increased the complexity of the data distribution. This makes the classification problem harder to solve than the original one considering the entire events as training examples. All of this, again, results in the degeneration of the classification models.

### B. The improved detection pipeline with a verification step

In order to mitigate the above-discussed shortcoming and take full advantage of high-quality classification models for the detection task, we propose to employ them in a completely different manner. The idea is that we augment a detection system with a verification step where a high-quality classification model will be applied to verify the detected event hypotheses as in Fig. 1. At this step, an event hypothesis with a class label $c_{detected}$ outputted by the detection system will be rejected when it is classified with a mismatched label $c_{classified} \neq c_{detected}$ by the classifier. Eventually, instead of making hard decisions early on, the false positive hypotheses outputted by a detection system will be rejected by the verification step. As a result, the detection precision will be enhanced, leading to improvements in overall detection performance.

The rationale behind the proposed pipeline originates from the difference between the classification and detection tasks as discussed above. Since the detection task relies on local features, the wrongly detected events are usually difficult ones whose local features are not reliable and cause wrong detections. However, after the detection step, we obtain the estimated boundary of the detected events and therefore, have access to their more or less global contexts. As a result, the mismatch between training and testing data is mitigated and an event classifier is expected to perform well. It should be also noted that we achieve the goal of false positive reduction by checking the consistency between the detection labels and the classification ones. Our approach, therefore, differs from that in [21] which relied on cascading classifiers.

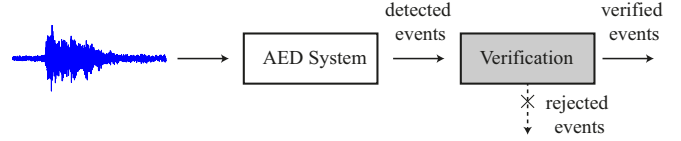

Fig. 1. The improved audio event detection pipeline with the verification step for false positive reduction.

## III. THE EMPLOYED EVENT DETECTORS AND CLASSIFIERS

In order to verify the effectiveness of the proposed detection pipeline, we implement three event detection systems and five different event classifiers and study all possible combinations of them in the proposed pipeline. These detectors are typical ones complying with three common detection schemes: detection-by-classification, ASR-based, and regression. The employed classifiers are chosen to be diverse enough, ranging from the standard bag-of-words model to the one recently reported state-of-the-art performance.

### A. Audio event detectors

**Detection-by-classification.** This system conforms to the common detection-by-classification scheme. It uses a sliding window of one second and a shift of 100 ms on audio signals for detection. The detection task is accomplished by two RBF-kernel SVM classifiers: one for event/background classification and the other for subsequent event classification. For representation, each one-second segment is decomposed into 25 ms frames with an overlap of 50%. A set of 60 features as used in [16] is then extracted for a frame: 16 log-frequency filter bank coefficients, their first and second derivatives, zero-crossing rate, short time energy, four subband energies, spectral flux calculated for each subband, spectral centroid, and spectral bandwidth. In turn, a global feature vector which consists of mean and standard deviation of frame-wise feature vectors is used to represent each one-second segment. Furthermore, a median filter of size 17 is applied on the label sequences to eliminate too short silences or non-silences [16]. Finally, an event hypothesis is excluded if its length is less than the minimum length of training instances.

**ASR-based.** This system adheres to the ASR framework. The audio signals are divided into short 20 ms audio frames with a hop size of 10 ms as commonly used for speech. Each frame is represented by twelve Mel-Frequency Cepstral Coefficients (MFCCs). Each event is described by a three-state HMM. All of the HMMs have a left-to-right topology and use output probability densities represented by means of 128 Gaussian components with diagonal covariance matrices. HMM training was accomplished through the Baum-Welch training procedure. Finally, the optimum event sequence is obtained by the Viterbi decoding algorithm.

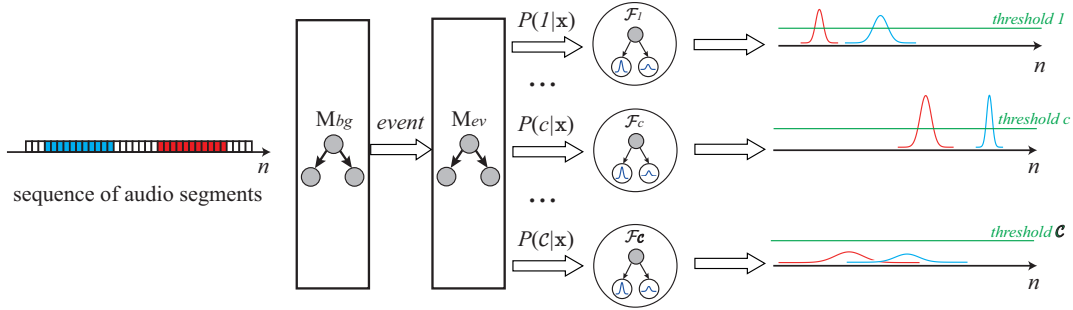

Fig. 2. The audio event detection system based on the regression approach.

**Regression.** This system follows the regression approach which was recently proposed for event detection and demonstrates state-of-the-art results [4], [13]. We show the system pipeline in Fig. 2 to ease the explanation not only for this detector itself but also the event classifiers (i.e. BoR and HoDW in Section III-B) which are derived from this pipeline. The audio signals are decomposed into 100 ms long segments with an overlap of 90 ms. We utilize the feature set used in the detection-by-classification scheme to represent each segment. Out of $\mathcal{C}$ event categories of interest, we trained a regression forest $\mathcal{F}^c$ for each category $c \in \{1, \ldots, \mathcal{C}\}$. The binary segment-wise classifier $\mathcal{M}_{bg}$ is used for event/background classification. The event segments are subsequently classified by the multi-class event classifier $\mathcal{M}_{ev}$. The segments classified as class $c$ will then be presented to regression forests $\mathcal{F}^c$ for event onset and offset estimation. In addition, the estimation is weighted by the classification probabilities obtained from $\mathcal{M}_{ev}$. Class-wise detection thresholds searched by cross-validation are eventually applied to the estimation scores to determine event onset and offset positions. Both $\mathcal{M}_{bg}$ and $\mathcal{M}_{ev}$ were trained using random forest classification [22] with 200 trees each. The regression forests were trained with the random forest regression algorithm as in [4], [13].

### B. Audio event classifiers

We implemented the following event classifiers to play the role of the verifier in the proposed detection pipeline.

**Bag-of-words (BoW).** BoW models have been widely used for audio event classification [23], [17]. Using this model, an audio event is represented by a histogram of codebook entries. The isolated events were divided into 50 ms long segments each of which is represented by the feature set used in the detection-by-classification detector. We used $k$-means for codebook learning. The entries were obtained as the cluster centroids, and codebook matching was based on Euclidean distance. We varied the codebook size in the set of $\{50, 100, \ldots, 250\}$. The system with the best performance was kept.

**Pyramid bag-of-words (PBoW).** As an improvement of the BoW models, we extracted and combined BoW descriptors on different pyramid levels [24] to encode the temporal structure of the audio events as in [17]. We exploited $\{2, 3, 4\}$ pyramid levels and the best one was retained.

**Bank-of-regressors (BoR).** This system utilizes the class-specific regression forests $\mathcal{F}^c$ in the regression-based detector in Fig. 2 and stacks them into a bank for feature extraction as in [25]. An audio event is then transformed into a compact descriptor $\phi = [\phi_1, \ldots, \phi_C]^T \in \mathbb{R}_+^C$ where $\phi_c$ is the mean of maximum onset and offset estimation scores outputted by the regressor $\mathcal{F}^c$. Each entry $\phi_c$ was normalized to $\frac{\phi_c}{\max_{\phi_c}}$, where $\max_{\phi_c}$ is the maximum value of $\phi_c$ in the training events. This descriptor is compact since its dimensionality is equal to the number of target event categories. Intuitively, the descriptor measures how the audio event aligns to the temporal configurations of different event categories modeled by the regressors.

**Histogram of discriminative words (HoDW).** This classifier also utilizes a component of the regression-based detector, the event classifier $\mathcal{M}_{ev}$, in Fig. 2 for feature extraction. An event consisting of $N$ audio segments $(\mathbf{x}_n; n = 1, \ldots, N)$ is represented by the descriptor $\varphi = [\varphi_1, \ldots, \varphi_C]^T \in \mathbb{R}_+^C$ where:

$$\varphi_c = \sum_{n=1}^{N} \frac{1}{N} P(c \mid \mathbf{x}_n, \mathcal{M}_{ev}). \quad (1)$$

In (1), $P(c \mid \mathbf{x}_n, \mathcal{M}_{ev})$ is the probability that the segment $\mathbf{x}_n$ is classified as class $c$ by the classifier $\mathcal{M}_{ev}$. These features can be thought of being a discriminative and compact variant of the BoW models where $\mathcal{M}_{ev}$ plays the role of the discriminative codebook matcher. In addition, opposed to the BoR descriptor, this one is unstructured and expected to be good for weakly structured events.

**BoR+HoDW.** Since BoR and HoDW are expected to be good for event types which expose strong and weak temporal structures, respectively, it is reasonable to combine both types of descriptors to take advantage of their strengths. In addition, this is convenient since they are derived from the same pipeline in Fig. 2. We combine two descriptors using an extended Gaussian kernel [26]:

$$K(e_i, e_j) = \exp\left(-\sum_{k \in \{\phi, \varphi\}} \frac{1}{A^k} D(e_i^k, e_j^k)\right). \quad (2)$$

where $D(e_i^k, e_j^k)$ is the $\chi^2$ distance between the audio events $e_i$ and $e_j$ with respect to the $k$-th feature channel. $A^k$ is the mean value of the $\chi^2$ distances between the training samples for the $k$-th channel.

TABLE I
OVERALL CLASSIFICATION ACCURACIES (%) OBTAINED BY DIFFERENT CLASSIFICATION SYSTEMS.

| BoW | PBoW | HoDW | BoR | BoR+HoDW |
|---|---|---|---|---|
| 97.3 | 96.6 | 97.8 | 98.4 | 98.9 |

All final classifiers were trained using one-vs-one SVMs and the hyperparameters were tuned via 10-fold cross-validation. The $\chi^2$ kernel was used for the BoW, PBoW, BoR, and HoDW classifiers whereas the kernel defined in (2) was used for the BoR+HoDW one.

## IV. EXPERIMENTS

### A. Experiment setup

*1) Dataset:* We conducted experiments on the ITC-Irst dataset [2]. It consists of twelve recording sessions with a total duration of 1.7 hours. There are 16 semantic event categories with approximately 50 events recorded for most of the categories. To be consent with the CLEAR 2006 challenge [2] and previous works [4], [13], we target twelve classes for evaluation: door knock (kn), door slam (ds), steps (st), chair moving (cm), spoon cup jingle (cl), paper wrapping (pw), key jingle (kj), keyboard typing (kt), phone ring (pr), applause (ap), cough (co), laugh (la). The rest are considered as background. We used nine recording sessions for training and three remaining sessions for testing. Only one channel named *TABLE_1* was used.

*2) Evaluation metrics:* We evaluate the performance of the detection systems using *F1-score* metric. For completeness, we also report the performance of the employed event classifiers in terms of overall classification accuracy. Note that due to the randomness of the system in Fig. 2, the experiments related to the BoR, HoDW, and BoR+HoDW classifiers as well as the regression-based detector were repeated five times and the average performance is reported.

*3) Experimental results:* The classification performances achieved by different systems are summarized in Table I. Both the BoW and PBoW classifiers yield best performance with a codebook size of 200. A pyramid level of 2 is optimal for the PBoW classifier. Between the BoW and HoDW classifiers which do not take into account the temporal structure of the events, the latter one saw an absolute improvement of 0.5% in classification accuracy, thanks to the discriminative codebook. Likewise for the classifiers using structural features, BoR outperforms PBoW by 1.8% absolute. This result explains that the BoR descriptor appears to model the temporal information of the events more efficiently than the PBoW one. Finally, as expected, integration of both structural and unstructural descriptors, i.e. BoR and HoDW, in the BoR+HoDW system boosts the classification accuracy by 0.5% absolute compared to that of the BoR classifier.

The overall detection results of the employed detection systems, both with and without verification, are shown in Table II. Without verification, the regression-based detection system obtains 93.1% in terms of F1-score which significantly outperforms the detection-by-classification and ASR-based detectors by 9.4% and 8.7% absolute, respectively.

The detection results with verification demonstrate the effectiveness of the proposed detection pipeline. Overall, the verification with all classifiers leads to improvements in overall detection performance for all detection systems. The average F1-score gains of 6.7%, 1.7%, and 0.9% are seen for detection-by-classification, ASR-based, and regression detection systems, respectively. The principle is that the verifiers reject the false positives, increasing the precision at the cost of decreasing the recall. The reduction in recall is due to accidental rejection of true positive hypotheses by the classifiers. This side effect is explainable since the classifiers are not perfect and the segmentation errors of the detected hypotheses cause a certain degree of mismatch between training and test data. However, the recall drops are sustainably smaller than the precision gains, which leads to overall improvements in F1-score. In addition, the gains are varied depending on detector-classifier combination. For the detection-by-classification and ASR-based detectors, the F1-score gains achieved by verification with all five classifiers are almost the same, i.e. the standard BoW model is as good as the best BoR+HoDW one in this sense. However, for the regression-based detector, the gains obtained by verification with the BoW and PBoW classifiers are subtle (0.0% and 0.3%, respectively) while those with the HoDW, BoR, and BoR+HoDW classifiers are more significant (1.2%, 1.5%, and 1.4%, respectively). Furthermore, these classifiers offer a unique advantage that using them for verification purposes is much more convenient than the other two since they utilize the existing components of the regression-based detector for feature extraction.

The experimental results also help to reveal the behavior of different detection systems. The detection-by-classification system retains a lot of false positive hypotheses which are explained by its high recall (87.7%) but low precision (80.1%). In contrast, the ASR-based system is much more conservative, maintaining relatively small number of hypotheses explaining its high precision (87.4%) but low recall (81.5%). Consequently, the verification step is able to reject a lot more false positive hypotheses of the detection-by-classification detector to boost the precision significantly (13.8% on average over all five classifiers) whereas the precision gain of the ASR-based system is more subtle (6.0% on average over all five classifiers). Regarding the regression-based detector, even though it obtains much higher precision and recall (i.e. the retained hypotheses are of higher fidelity) than those of other two detectors, the verification step can further help to reject false alarms and yield significant gain in the precision, for example 3.7% with the BoR classifier.

## V. CONCLUSIONS

In summary, we studied what make audio event detection harder than classification. We then leverage this observation to study an important aspect of audio event detection systems on continuous streams, namely false positive reduction. An improved detection pipeline is proposed by appending a verification step to augment a detection system where an event classifier postprocesses the outputted event hypotheses and rejects the false positive ones. Three detection systems based on different detection schemes were implemented and coupled with various event classifiers which play the role of the verifier. The detection results with the verification step show consistent improvements on overall detection performance over all detector-classifier combinations. These results demonstrate the effectiveness of the proposed detection pipeline.

TABLE II
OVERALL DETECTION RESULTS OF DIFFERENT DETECTION SYSTEMS WITH AND WITHOUT VERIFICATION. FOR THOSE WITH VERIFICATION, THE
ABSOLUTE PERFORMANCE GAINS AND LOSSES COMPARED TO THOSE WITHOUT VERIFICATION ARE HIGHLIGHTED IN BLUE AND RED, RESPECTIVELY.

|  |  | Detection-by-classification | | | ASR-based | | | Regression | | |
|---|---|---|---|---|---|---|---|---|---|---|
|  |  | F1-score | precision | recall | F1-score | precision | recall | F1-score | precision | recall |
| w/o verification | | 83.7 | 80.1 | 87.7 | 84.4 | 87.4 | 81.5 | 93.1 | 93.6 | 92.7 |
| w/ verification | BoW | 90.5 ↑6.8 | 93.6 ↑13.5 | 87.7 ↓0.0 | 86.2 ↑1.8 | 94.3 ↑6.9 | 79.5 ↓2.0 | 93.0 ↑0.0 | 96.4 ↑2.8 | 89.9 ↓2.8 |
|  | PBoW | 90.2 ↑6.5 | 93.0 ↑12.9 | 87.7 ↓0.0 | 86.1 ↑1.7 | 95.0 ↑7.6 | 78.8 ↓2.7 | 93.4 ↑0.3 | 96.4 ↑2.8 | 90.7 ↓2.0 |
|  | HoDW | 90.6 ↑6.9 | 93.8 ↑13.7 | 87.7 ↓0.0 | 86.1 ↑1.7 | 91.9 ↑4.5 | 81.0 ↓0.5 | 94.3 ↑1.2 | 96.6 ↑3.0 | 92.1 ↓0.6 |
|  | BoR | 90.5 ↑6.8 | 95.2 ↑15.1 | 86.3 ↓1.4 | 85.7 ↑1.3 | 92.9 ↑5.5 | 79.5 ↓2.0 | 94.6 ↑1.5 | 97.3 ↑3.7 | 92.1 ↓0.6 |
|  | BoR+HoDW | 90.4 ↑6.7 | 93.8 ↑13.7 | 87.3 ↓0.4 | 86.2 ↑1.8 | 92.8 ↑5.4 | 80.4 ↓1.1 | 94.5 ↑1.4 | 97.0 ↑3.4 | 92.1 ↓0.6 |

ACKNOWLEDGMENT

This work was supported by the Graduate School for Computing in Medicine and Life Sciences funded by Germany's Excellence Initiative [DFG GSC 235/1].